\documentclass{article}

\usepackage{algpseudocode}
\usepackage{amsmath}
\usepackage{booktabs}
\usepackage{cite}
\usepackage{float}
\usepackage{graphicx}
\usepackage{subfig}
\usepackage{wrapfig}
\usepackage{hyperref}
\usepackage{microtype}
\usepackage{comment}
\usepackage{paralist}
\usepackage{times}

\hypersetup{
    unicode=true,
    pdftoolbar=true,
    pdfmenubar=true,
    pdffitwindow=false,
    pdfstartview={FitH},
    pdftitle={Network-on-Chip Firewall: Countering Defective and Malicious System-on-Chip Hardware},
    pdfauthor={Michael LeMay and Carl A. Gunter},
    pdfsubject={Network-on-Chip Security},
    pdfnewwindow=true,
    colorlinks=true,
    citecolor=black,
    filecolor=black,
    linkcolor=black,
    urlcolor=black
}

\floatstyle{ruled}
\newfloat{listing}{H}{lst}
\floatname{listing}{Listing}

\newcommand{\widepseudocode}[3]{
\begin{listing}[H]
  \begin{algorithmic}
    #3
  \end{algorithmic}
\caption{#2}
\label{pseudo:#1}
\end{listing}
}

\newcommand{\includefig}[3]{
\begin{figure}%
  \centering
    \includegraphics[width=#3\columnwidth]{#1}
  \caption{#2}
  \label{fig:#1}
\end{figure}
}

\newcommand*{\mathsymb}[1]{\expandafter \newcommand \csname
  ms#1\endcsname{\textit{#1}}}

\newcommand{\sys}{NoCF}
\newcommand{\etal}{et~al.}

\newtheorem{invariant}{Invariant}

\begin{document}

\title{Network-on-Chip Firewall: Countering Defective and Malicious System-on-Chip Hardware}

\author{
Michael~LeMay\thanks{M. LeMay was with the University of Illinois at Urbana-Champaign while performing the work described herein, but he was employed by Intel Corporation at the time of submission. The views expressed are those of the authors only.}, Carl~A.~Gunter\\University of Illinois at Urbana-Champaign
}

\maketitle

\begin{abstract}
Mobile devices are in roles where the integrity and confidentiality of their apps and data are of paramount importance.  They usually contain a System-on-Chip (SoC), which integrates microprocessors and peripheral Intellectual Property (IP) connected by a Network-on-Chip (NoC).  Malicious IP or software could compromise critical data.  Some types of attacks can be blocked by controlling data transfers on the NoC using Memory Management Units (MMUs) and other access control mechanisms.  However, commodity processors do not provide strong assurances regarding the correctness of such mechanisms, and it is challenging to verify that all access control mechanisms in the system are correctly configured.  We propose a NoC Firewall (NoCF) that provides a single locus of control and is amenable to formal analysis.  We demonstrate an initial analysis of its ability to resist malformed NoC commands, which we believe is the first effort to detect vulnerabilities that arise from NoC protocol violations perpetrated by erroneous or malicious IP.
\end{abstract}

\section{Introduction}
\label{sect:intro}

Personally administered mobile devices are being used or considered for banking, business, military, and healthcare applications where integrity and confidentiality are of paramount importance.  The practice of dedicating an entire centrally administered phone to each of these apps is being abandoned in favor of granting access to enterprise data from personal devices as workers demand the sophistication available in the latest consumer mobile devices~\cite{cheng_so_2011}.

Security weaknesses of popular smartphone OSes have motivated the use of isolation mechanisms on devices entrusted with critical data, including hypervisors that operate at a lower level within the system~\cite{nachenberg_window_2011}.  For example, hypervisors can isolate a personal instance of Android from a sensitive instance of Android, where both instances run simultaneously within Virtual Machines (VMs) on a single physical device.  However, virtualized and non-virtualized systems both rely on the correctness of various hardware structures to enforce the memory access control policies that the system software specifies to enforce isolation.

Mobile devices are usually based on a System-on-Chip (SoC) containing microprocessor cores and peripherals connected by a Network-on-Chip (NoC).  Each component on the SoC is a piece of Intellectual Property (IP).  SoC IP may be malicious intrinsically at the hardware level, or it may be used to perform an attack orchestrated by software, and such IP may lead to compromises of critical data.  Such attacks would involve data transfers over the NoC.  Memory Management Units (MMUs) and IO-MMUs can potentially prevent such attacks.

Commodity processors do not provide strong assurances that they correctly enforce memory access controls, but recent trends in system design may make it feasible to provide such assurances using enhanced hardware that is amenable to formal analysis.  In this paper, we propose the hardware-based \textit{Network-on-Chip Firewall (\sys{})} that we developed using a functional hardware description language, Bluespec.  Although Bluespec has semantics based on term-rewriting systems, those semantics also reflect characteristics of hardware~\cite{_bluespec_overview_2006}.  We developed an embedding of Bluespec into Maude, which is a language and set of tools for analyzing term-rewriting systems.  At a high level, term-rewriting systems involve the use of atomic rules to transform the state of a system.  We know of no elegant way to directly express the hardware-specific aspects of Bluespec in a Maude term-rewriting theory, so we used Maude strategies to control the sequencing between rules in the theories to match the hardware semantics.  We then used our model to detect attacks that violate NoC port specifications, which have previously received little attention.

A lightweight processor core is dedicated to specifying the NoCF policy using a set of \textit{policy configuration interconnects} to interposers, which provides a single locus of control.  It also permits NoCF to be applied to NoCs lacking access to memory, avoids the need to reserve system memory for storing policies when that memory is available, and simplifies the internal logic of the interposers.  The policy can be pre-installed or specified dynamically by some entity such as a hypervisor within the system.  The interposers and associated policies are distributed to accommodate large NoCs.

We use a triple-core FPGA prototype running two isolated instances of Linux to demonstrate that \sys{} introduces negligible performance overhead and can be expected to only slightly increase hardware resource utilization when enforcing coarse-grained memory access control policies.

To demonstrate one type of attack that can be blocked by NoCF, we construct malicious IP analogous to a Graphics Processing Unit (GPU) and show how it can be instructed to install a network keylogger by any app that simply has the ability to display graphics.  This attack could be used to achieve realistic, malicious objectives.  For example, a government seeking to oppress dissidents could convince them to view an image through a web browser or social networking app and subsequently record all of their keystrokes.

Our contributions include:
\begin{compactitem}
 \item An efficient, compact \sys{} interposer design that is amenable to formal analysis and provides a single locus of control.
 \item An embedding of Bluespec into the Maude modeling language.
 \item Use of formal techniques to discover a new attack.
 \item A triple-core FPGA prototype that simultaneously runs two completely isolated, off-the-shelf instances of Linux with no hypervisor present on the cores or attached to the NoCs hosting Linux at runtime.
\end{compactitem}

The rest of this paper is organized as follows.  \S{}\ref{sect:bg} provides background on sensitive apps for mobile device and on SoC technology.  \S{}\ref{sect:design} discusses the design of \sys{}.  \S{}\ref{sect:eval} evaluates a \sys{} prototype.  \S{}\ref{sect:formal} formally analyzes that prototype.  \S{}\ref{sect:discussion} discusses other potential uses for \sys{} and ways to improve it.  \S{}\ref{sect:related} discusses related work.  \S{}\ref{sect:conclusion} concludes the paper.

\section{Background}
\label{sect:bg}

\subsection{Mobile Device Applications}
\label{sect:bg:mobile-apps}

Sensitive information from various sectors is commonly accessed using mobile devices~\cite{power_mobility_2011}.
The particular usages of concern are rapidly evolving and somewhat unpredictable.  For example, mobile banking apps have quickly gained widespread acceptance and run alongside other apps.  They have capabilities such as supporting check deposits based on camera images that could permit attackers to insert themselves into critical banking transactions if they compromised the host device.  We highlight some particularly interesting apps in this section.

The military has considered supporting apps on commercial smartphones that soldiers already possess and that are used to perform mainstream smartphone tasks such as accessing the Internet, or that the military supplies to them for dedicated use.  Those apps could include support for tracking other soldiers using GPS, sharing smartphone and Unmanned Aerial Vehicle (UAV) sensor data~\cite{erwin_smartphones-for-soldiers_2011}, piloting UAVs~\cite{hennigan_taking_2011}, entering intelligence information~\cite{_is_2011,zakaria_not_2011}, identifying suspect vehicles and people, managing medical records~\cite{johnson_army_2011}, and providing realtime guidance to paratroopers using a mobile device during descent and after landing~\cite{ackerman_new_2011}.

Mobile devices in the hands of both patients and clinicians are broadly applicable to healthcare.  They have been used to perform ultrasound imaging, screen for skin cancer, monitor sleep patterns and other physical activity, plan surgeries, and perform cytometry (counting and examining cells in body fluids)~\cite{schultz_fda_2011,butcher_check_2011,lewis_hallux_2011,stomp_cell_2011,_up_2011}.
It is possible that iPads will be able to replace many pieces of equipment in operating rooms~\cite{wodajo_future_2011}.
Electronic Health Records (EHRs) can be accessed using native apps or web apps.  The Department of Veterans Affairs permits its employees to view and potentially store sensitive data on personal Apple devices~\cite{mosquera_data_2011,chopra_smart_2011}.
Smartphones can also be used in a clinical setting to send and receive phone calls and text messages containing patient data~\cite{duffy_doximity_2011,eytan_group_2011}.

An analysis of sharing of compromised medical data on peer-to-peer networks indicates that attackers are already targeting such data~\cite{johnson_data_2009}.  An attacker may be able to embarrass the patient using information stolen from their medical records, or perhaps use it to devise a strategy for harming the patient.  For example, the attacker could learn the exact model information of the patients' Implantable Cardiac Defibrillator that could be used to identify feasible attacks to send malicious commands to the ICD to harm the patient.  However, such attacks would have other prerequisites, such as equipment for attacking the ICD and close physical proximity~\cite{halperin_pacemakers_2008}.  In contrast, attackers could fairly immediately harm a diabetic patient by remotely attacking an app that displays the level of glucose in their blood, as measured by a phone-compatible blood glucose monitor.  The data from those apps may influence inputs to insulin delivery devices.  Paul~\etal{} discuss the functionality and criticality of similar devices at length~\cite{nathanael_paul_review_2011}.

Smartphone apps exist that can access SCADA devices on the electric power grid~\cite{bob_lockhart_utility_2011}.  Some SCADA devices are capable of changing the physical flow of electricity, so it is critical that apps used to control such devices be isolated from malware.

In the future, SoCs may also be used to power cloud servers, and could thus benefit from the protections provided by \sys{} to separate workloads from different clients that are assigned to run on a single SoC.

\subsection{System-on-Chip Technology}
\label{sect:bg:soc-tech}
An SoC comprises a collection of IP that is processed to result in a final silicon chip that performs functions that used to be spread across several chips~\cite{_soc_2011}.  For example, the OMAP44x series from Texas Instruments integrates the following hardware plus more onto a single chip: two ARM cores, graphics accelerator, HDMI controller, timers, interrupt controller, boot ROM, cryptographic accelerators, USB controller, FLASH and SDRAM memory controllers, and a serial port controller.  This affects the design and location of the interconnects between those components.  Data transfers within a chip are fast and essentially error-free compared to off-chip transfers.

The SoC development strategy has the advantage of feasibly permitting the introduction of security-enhanced hardware in a particular SoC vendor's design without requiring that it be included in all designs that are based on the microprocessor IP contained within the SoC.  Another factor that increases the feasibility of incorporating such technology is the low incremental cost of adding functionality to an SoC, since it does not require a separate chip.  This is particularly true if the functionality can make use of the increasingly-plentiful dark silicon, which is silicon that can only be activated when other silicon is deactivated, due to thermal constraints~\cite{esmaeilzadeh_dark_2011}.  It is increasing due to the additional heat generated by modern silicon manufacturing processes that use smaller transistors, since the heat dissipation capacity of silicon processors has not improved at a similar pace.  \sys{} is entirely contained within the SoC, and parts of it may be able to use dark silicon.

Each piece of IP on an SoC can be provided by an organization within the SoC vendor or by an external organization.  SoCs commonly contain IP originating from up to hundreds of people in multiple organizations and spread across multiple countries~\cite{villasenor_ensuring_2011}.  It is difficult to ensure that all of the IP is high-quality, let alone trustworthy~\cite{goering_panelists_2011,butler_managing_2011}.
The general trend is towards large SoC vendors acquiring companies to bring IP development in-house~\cite{sn2011}.  However, even in-house IP may provide varying levels of assurance depending on the particular development practices and teams involved and the exact nature of the IP in question.  For example, a cutting-edge, complex GPU may reasonably be expected to exhibit more errors than a relatively simple Wi-Fi controller that has been in use for several years.  Memory Management Units (MMUs) and IO-MMUs are commonly used to restrict the accesses from IP, which can constrain the effects of erroneous or malicious IP.  Thus, errors that can permit memory access control policies to be violated are the most concerning.

An MMU is a component within a processor core that enforces memory access control policies specified in the form of page tables that are stored in main memory.  Some SoCs incorporate IO-MMUs that similarly restrict and redirect peripheral master IP NoC data transfers.  A page table contains entries that are indexed by part of a virtual address and specify a physical address to which the virtual address should be mapped, permissions that restrict the accesses performed using virtual addresses mapped by that entry, whether the processor must be in privileged (supervisor) mode when the access is performed, and auxiliary data.  Page tables are often arranged hierarchically in memory, necessitating multiple memory accesses to map a particular virtual address.  To reduce the expense incurred by page table lookups, the MMU contains a Translation Lookaside Buffer (TLB) that caches page table entries in very fast memory inside the MMU.  Each isolated software component (such as a process or VM) is typically assigned a dedicated page table.  By only mapping a particular region of physical memory in one of the component's page tables, that memory is protected from accesses by other components.  The relatively high complexity of modern MMUs and IO-MMUs increases the likelihood of errors that undermine their access control assurances~\cite{gotze_survey_2011}.  \sys{} is much less complex and can constrain an attack leveraging a vulnerable MMU or IO-MMU.

It could be preferable to formally verify existing MMUs and IO-MMUs rather than devising new protection mechanisms.  However, it is challenging to formally verify MMUs and IO-MMUs.  Formal verification techniques can prove the absence of design errors within commercial processor cores, but they currently only provide a good return-on-investment when used instead to detect errors~\cite{arditi_formal_2010}.  To the best of our knowledge, MMUs have only been formally verified in experimental processors~\cite{dalinger_formal_2006,schubert_formal_1992}.  The policy data for MMUs and IO-MMUs is itself protected by them, so it is likely to be more challenging to verify that the policy is trustworthy compared to the \sys{} policy implemented on an isolated core.  Finally, the MMU is a central part of each processor core with many interfaces to other parts of the core, complicating analysis.  We have not formally verified \sys{} either, but we demonstrate how to develop a model of it that is amenable to formal analysis.  This is a non-trivial precondition for formal verification.

Individual pieces of IP communicate using one or more NoCs within a single SoC.  A NoC is not simply a scaled-down network comparable to, e.g. an Ethernet LAN.  Networks for large systems, such as LANs, have traditionally been connection-oriented, predominantly relying on protocols such as TCP/IP.  Networks for small systems, such as NoCs, have traditionally lacked support for persistent connections.  Older SoC designs relied on buses, which are subtly distinct from NoCs.  For our purposes, it is not necessary to distinguish between buses and NoCs.  We are concerned primarily with their external ports, which are common between both types of interconnects.  Slave devices accessible over a NoC are assigned ranges of physical addresses, so memory access controls like those in \sys{} can also be used to control access to devices besides actual RAM controllers.

Malicious hardware can be inserted at many points within the SoC design and manufacturing process and can exhibit a variety of behaviors to undermine the security assurances of the system~\cite{beaumont_hardware_2011}.  For example, malicious hardware can watch for a particular sequence of bytes on a data bus and then trigger the MMU to start ignoring CPU privilege levels when implementing memory access control policies, so that unprivileged code can access privileged memory regions~\cite{king_designing_2008}.  Various countermeasures have been devised to contain or otherwise disrupt malicious hardware, but some of them require access to the source code of the IP and many of them have various functional drawbacks~\cite{beaumont_hardware_2011}.  We discuss how \sys{} can help defend against malicious IP.  Some of those previously developed techniques may be complementary to ours when they are feasible.

The protection mechanisms that we propose are inserted between the NoC and the IP, and they do not necessitate changes to individual IP blocks. Thus, \sys{} could be added quite late in the design process for an SoC, after the main functionality of the SoC has been implemented.

\section{Design}
\label{sect:design}

\subsection{Threat Model}
\label{sect:design:threat}

Software running on a particular core is assumed to be arbitrarily malicious and must be prevented from compromising the confidentiality, integrity, and availability of software on other cores.  The system software that configures \sys{} must correctly specify a policy to enforce isolation between the cores.  Recent work on minimizing the Trusted Computing Base (TCB) of hypervisors and formally verifying them may be helpful in satisfying this requirement~\cite{klein_sel4:_2009,szefer_eliminating_2011}.

Our concern in this paper is that isolation between cores that are protected in this manner could potentially be compromised by misbehaving IP.  We now define the types of compromises we seek to prevent:
\begin{compactenum}
 \item \textit{Confidentiality:} Some misbehaving IP may construct an unauthorized information flow from some other target IP transferring data that the misbehaving IP or the VM controlling it is not authorized to receive.  This flow may be constructed with or without the cooperation of the target IP.  The misbehaving IP may have authorization to access a portion of the target IP, but not the portion containing the confidential data, e.g. in the case of a shared memory controller.

 \item \textit{Integrity:} Some misbehaving IP may unilaterally construct an unauthorized information flow to some other target IP transferring data that masquerades as data from a different originator.  For example, it may modify executable code stored in shared memory that is used by a VM, or modify medical sensor data that is displayed to a user.

 \item \textit{Availability:} Resource sharing is an intrinsic characteristic of SoCs, so there is the possibility that misbehaving IP may interfere with other IP using those shared resources.  For example, the misbehaving IP could flood the NoC with requests to monopolize the NoC.
\end{compactenum}

IP can manipulate wires that form its NoC port in an arbitrary manner.  The IP might not respect the port clock and can perform intra-clock cycle wire manipulations.  The IP might also violate the protocol specification for the port.

Since \sys{} performs address-based access control, we trust the NoC fabric to selectively and accurately route requests and responses to and from the appropriate IP, so that other IP cores do not have the opportunity to eavesdrop or interfere.  Currently-available fabrics may in fact not be trustworthy and resistant to attacks, which should be the subject of future research.

We trust slave devices to correctly process requests.  For example, the memory controller must properly process addresses that it receives to enforce policies that grant different IP cores access to different regions of a memory accessible through a single shared memory controller.  Verification of such devices is an important, orthogonal area of research.

\begin{figure}
 \centering

 \subfloat[Unaltered, hypervisor-based system.]{\label{fig:isolation:orig-tcb}\includegraphics[width=0.4\columnwidth]{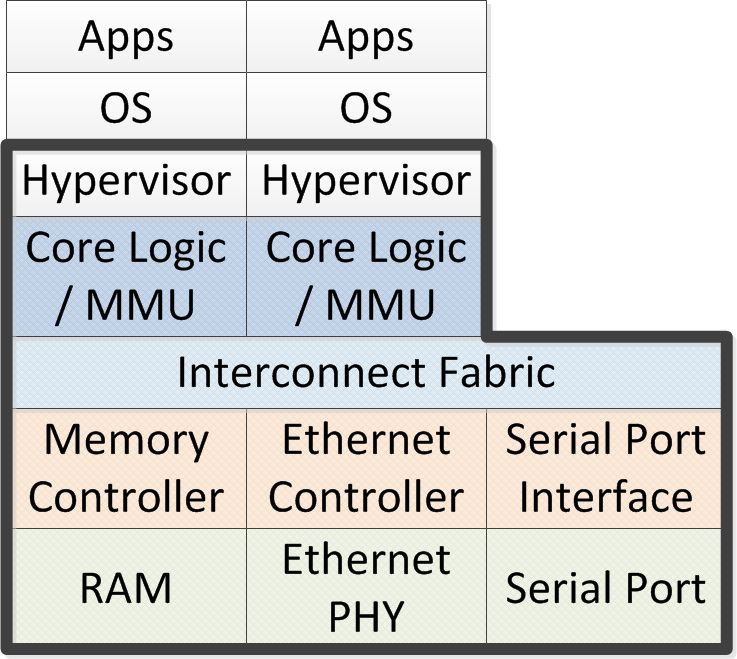}}
\qquad %
 \subfloat[Similar system protected by \sys{}.]{\label{fig:isolation:new-tcb}\includegraphics[width=0.4\columnwidth]{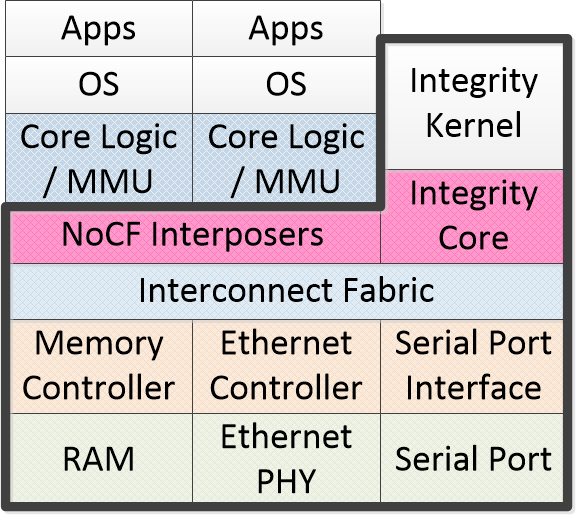}}

 \caption{Comparison of TCBs, which are within the thick lines.  Colored areas depict layers of hardware.}
\end{figure}

Covert channels are more prevalent between components that have a high degree of resource sharing, such as between software that shares a processor cache.  Thus, \sys{} provides tools to limit covert channels by restricting resource sharing.  However, we do not attempt to eliminate covert channels in this work.

A mobile device may be affected by radiation and other environmental influences that cause unpredictable modifications of internal system state.  A variety of approaches can handle such events and are complementary to our effort to handle misbehaviors that arise from the design of the device~\cite{mukherjee_soft_2005}.

System software could be maliciously altered.  Trusted computing techniques can defeat such attacks by ensuring that only specific system software is allowed to execute~\cite{arbaugh_secure_1997}.  We focus on techniques whereby the SoC vendor can constrain untrustworthy IP in its chip designs. Software security and hardware tamper-resistance techniques can further improve assurances of overall system security.

\subsection{Core-Based Isolation}
\label{sect:design:isolation}

Assigning software components to separate cores eliminates vulnerabilities stemming from shared resources such as registers and L1 caches.  Regulating their activities on NoCs with a dynamic policy addresses vulnerabilities resulting from sharing main memory or peripherals.  We initially focus on isolating complete OS instances on separate cores, since the memory access control policies required to accomplish that are straightforward and coarse-grained.  However, \sys{} could also be used to implement other types of policies.

The \sys{} policy either needs to be predetermined or defined by a hypervisor, like the hypervisor specifies MMU policies.  The policy will actually be maintained by an \textit{integrity kernel} that runs on a dedicated \textit{integrity core}, which will be discussed further below.  The effect that this has on the TCB of a system with minimal resource sharing, such as our prototype system that isolates two Linux instances, is depicted in Figure~1.  The TCB will vary depending on how the policy is defined, since any software that can influence the policy is part of the TCB.  In this example, the policy that was originally defined in a hypervisor is now defined in the integrity kernel, completely eliminating the hypervisor.

\sys{} provides a coarser and more trustworthy level of memory protection in addition to that of the MMU and IO-MMU.  These differing mechanisms can be used together to implement trade-offs between isolation assurances and costs stemming from an increased number of cores and related infrastructure.

The integrity core must be able to install policies in \sys{} interposers and must have sufficient connectivity to receive policy information from any other system entities that are permitted to influence policies, such as a hypervisor.  It may be possible to place the integrity kernel in firmware with no capability to communicate with the rest of the system, if a fixed resource allocation is desired.  On the other end of the spectrum of possible designs, the integrity core may have full access to main memory, so it can arbitrarily inspect and modify system state.  Alternately, it may have a narrow communication channel to a hypervisor.  Placing the integrity kernel on an isolated integrity core permits the pair of them to be analyzed separately from the rest of the system.  However, it is also possible to assign the role of integrity core to a main processor core to reduce hardware resource utilization, even if the core is running other code.

\subsection{\sys{} Interposers}
\label{sect:design:interposers}

\includefig{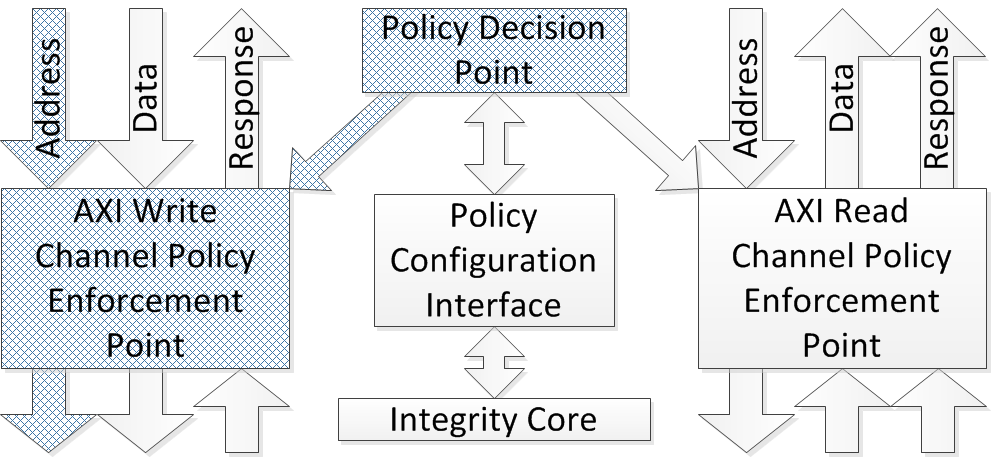}{Internal configuration of \sys{} interposer.  Each interposer contains all of these components.  Hatched regions are formally analyzed in \S{}\ref{sect:formal}.}{0.7}

We now discuss the design decisions underlying \sys{}.  We base our design on the widely-used AMBA AXI4 NoC port standard.  The rule format and storage mechanism of the Policy Decision Point (PDP) are loosely modeled after those of a TLB.  The PDP decides which accesses should be permitted so that a Policy Enforcement Point (PEP) can enforce those decisions.  Policy rules are inserted directly into the PDP using a policy configuration interconnect to an integrity core.  This reduces the TCB of the PDP relative to a possible alternate design that retrieves rules from memory like an MMU.  The integrity core is a dedicated, lightweight processor core that is isolated from the rest of the system to help protect it from attack.  The decisions from the PDP are enforced for each address request channel by that channel's PEP.

The AXI4 specification defines two matching port types.  The master port issues requests and the slave port responds to those requests.  Each pair of ports has two distinct channels, one for read requests and one for write requests.  This port architecture enables us to easily insert \sys{} interposers, each of which contains a PDP, an integrity core interface, and two PEPs, one for each channel.  Each interposer provides both a master and slave port so that it can be interposed between each IP master port and the NoC slave port that it connects to.  A single interposer is depicted in Figure~\ref{fig:internals}.

We evaluate our design in a prototype system containing two main processor cores in addition to the integrity core, plus a malicious GPU.  We now consider it as an example of how such a system is arranged, although many other system arrangements are possible.  Each main core has four AXI4 master ports.  They connect to two NoCs in the system, one of which solely provides access to the DDR3 main memory controller, while the other provides access to the other system peripherals.  Each main core has two ports connected to each NoC, one for instruction accesses and the other for data accesses.  The GPU has a single master port connected to the NoC with the main memory, along with a slave port connected to the peripheral NoC (not shown).  We depict this topology in Figure~\ref{fig:topology}.  One interposer is assigned to each of the ports between the master IP and the NoCs, with a corresponding policy configuration interconnect to the integrity core.  The depicted interconnect topology is slightly simplified compared to the one used in the commercial ARM Cortex-A9 MP processor, which shares an L2 cache between up to four cores.  Thus, it would be necessary in that processor to place interposers between the cores and the L2 cache controller and to trust that controller to implement memory addressing correctly.

\includefig{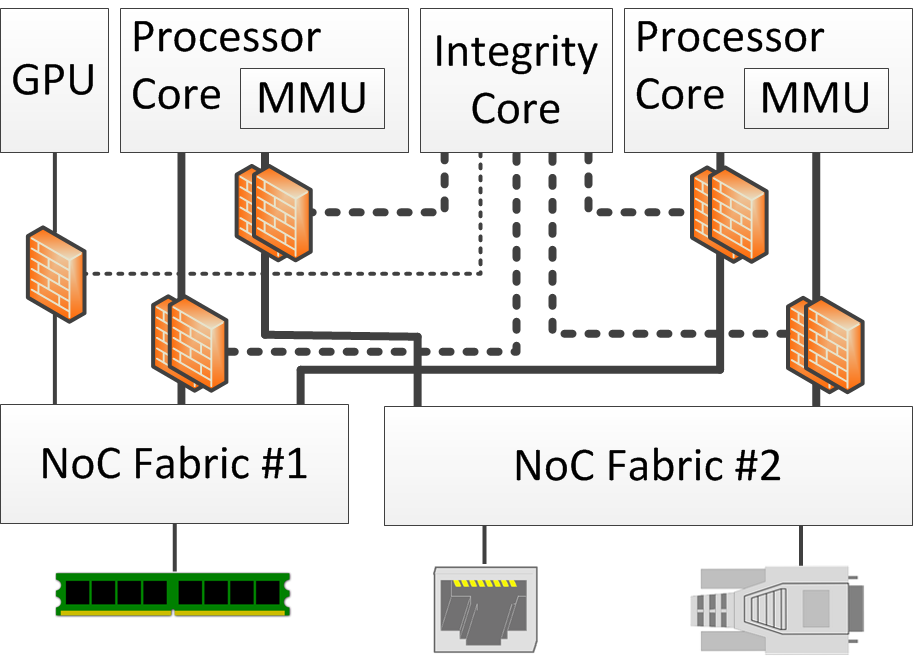}{System topology.  Each brick wall represents a \sys{} interposer on one NoC port.  Dashed lines denote policy configuration interconnects.  Solid lines denote NoC ports.  For the interconnects and ports, thin lines denote single items and thick lines denote pairs of items.}{0.55}

The distributed nature of the \sys{} interposers enables them to each use a policy tailored to the port being regulated and also concentrates the internal interfaces containing many wires between the PDP and PEPs in a small area of the chip while using an interface containing few wires to span the potentially long distance to the integrity core.  However, it may be useful in some cases to share a PDP between several interposers that are subject to a single policy.  That approach would reduce the number of policy configuration interconnects and the total PDP policy storage.  A more complex approach would be to support selectively-shared rules for separate interposers in a single PDP, which would still reduce interconnect logic and could provide some reduction in PDP storage.

Each policy rule specifies a region of memory to which read and/or write access is permitted.  A region is defined by a base address and a mask length specifier, which indicates the size of the region as one of a set of possible powers of two.  This type of policy can be implemented very efficiently in hardware and corresponds closely to the policies defined by conventional MMU page tables.

Address requests are regulated by the PEPs in cooperation with the PDP.  The PDP stores a fixed number of rules in its database.  The PDP checks the address in the request against all policy rules in parallel.  Whenever a request matches some rule that has the appropriate read or write permission bit set, it will be permitted to pass through the PEP.  Otherwise, the PDP sends an interrupt to the integrity core and also sends it information about the failing request.  It then blocks the request until the integrity core instructs it to resume.

The integrity core can modify the policy rules prior to issuing the resume command.  To modify policy rules, the integrity core sends commands over the policy configuration interconnect to insert a policy rule or flush all existing policy rules.  Other commands could be defined in the future.  When the interposer receives the resume command, it re-checks the request and either forwards it if it now matches some rule, or drops it and returns an error response to the master.  It could also do something more drastic, such as blocking the clock signal or power lines feeding the master that issued the bad request.

Addresses other than the one in the request may be accessed during the ensuing data transfer.  A variety of addressing modes are supported by AXI4 that permit access to many bytes in a burst of data transfers initiated by a single request.  The policy administrator must account for these complexities by ensuring that all bytes that can actually be accessed should be accessible.

It can be useful to physically separate a protection mechanism from the surrounding logic and constrain its interfaces to that logic so that it can be independently analyzed~\cite{HuffmireBWSKLNI08}.  This is possible in the case of \sys{}, since its only interfaces are the controlled NoC ports and the policy configuration interconnect.

\section{Evaluation}
\label{sect:eval}

\subsection{Prototype Implementation and Hypervisor Functionality}
\label{sect:eval:prototype}

We used a Xilinx ML605 evaluation board, which includes a Virtex-6 FPGA, to implement a prototype of \sys{}.  We use MicroBlaze architecture processor cores implemented by Xilinx, because they are well-supported by Xilinx tools and Linux.  The main cores allocate 16KiB to each instruction and data cache.  The integrity core is very lightweight, with no cache, MMU, or superfluous optional instructions.  It is equipped with a 16KiB block of on-chip RAM directly and exclusively connected to the instruction and data memory ports on the integrity core.  This RAM is thus inaccessible from the other cores.  This block is initialized from an integrity kernel firmware image when the FPGA is configured.  The PDP, integrity core interface, and PEP are all implemented in Bluespec to leverage its elegant semantics and concision, with interface logic to the rest of the hardware system written in Verilog and VHDL.

The prototype runs Linux 3.1.0-rc2 on both main cores, including support for a serial console from each core and exclusive Ethernet access from the first core.  We compiled the Linux kernel using two distinct configurations corresponding to the cores so that they use different regions of system memory and different sets of peripherals.  This means that no hypervisor beyond the integrity kernel is required, because the instances are completely separated.  The system images are loaded directly into RAM using a debugger.

The integrity kernel specifies a policy that constrains each Linux instance to the minimal memory regions that are required to grant access to the memory and peripherals allocated to the instance.  Attempts to access addresses outside of an instance's assigned memory regions cause the instance to crash with a bus error, which is the same behavior exhibited by a system with or without NoCF when an instance attempts to access non-existent physical memory.

The NoCF interposer is inserted at each slave port exported by the NoC IP, a total of nine ports.  The interposers each contain two policy rules and replace them in FIFO order, except that the interposers for data loads and stores to the peripherals contain four policy rules each, since they are configured to regulate fine-grained memory regions.  We evaluated two types of policy configuration interconnects with differing levels of resource usage.  One design creates a bi-directional pair of direct links between each interposer and the integrity core using Fast Simplex Link (FSL) IP provided by Xilinx.  Each FSL link only supports sending fixed-size packets of data in one direction.  The second design groups multiple interposers behind a pair of AXI slave interfaces, which then implement direct links to each interposer.  The NoCF interposer implementations for the various sizes of policy storage and interconnect types all comprise between 1494 and 1571 lines of Verilog generated by the Bluespec compiler from source files containing 518 and 512 lines for the FSL and AXI variants, respectively.  The integrity kernel firmware is compiled from 239 and 262 lines of C  (as measured by Wheeler's SLOCCount) for the FSL and AXI variants, respectively, excluding third-party libraries.  To demonstrate the simplicity of the kernel, we describe it using pseudocode in Appendix~\ref{app:integrity-kernel} and discuss how it compares to kernels that rely on MMUs or IO-MMUs for protection.

\subsection{Constraining a Malicious GPU}
\label{sect:eval:mal-ip}

A malicious GPU could perform powerful attacks, since it would have bus-master access and be accessible from all apps on popular mobile OSes.  Even an app that requests no explicit privileges can display graphics, so permission-based controls are useless to prevent GPU-based attacks.  Furthermore, almost all apps have a legitimate need to display graphics, so software protection mechanisms that analyze app behavior could not be expected to flag communications with the GPU as suspicious.

We constructed hardware IP that is analogous to a hypothetical malicious GPU.  It has both master and slave AXI4 interfaces.  In response to commands received on its slave interface, the IP reads data from a specified location in physical memory.  This is analogous to reading a framebuffer.  The IP inspects the least significant byte of each pixel at the beginning of the framebuffer.  This is a very basic form of steganographic encoding that only affects the value of a single color in the pixel, to reduce the chance of an alert user visually detecting the embedded data.  More effective steganographic techniques could easily be devised.  If those bytes have a specific ``trigger'' value, then the IP knows that part of the framebuffer contains a malicious command.  The trigger value is selected so that it is unlikely to appear in normal images.  The IP then continues reading steganographically-embedded data from the image and interprets it as a command to write arbitrary data embedded in the image to an arbitrary location in physical memory.

We developed a simple network keylogger to be injected using the malicious IP.  The target Linux system receives user input via a serial console, so the keylogger modifies the interrupt service routine for the serial port to invoke the main keylogger routine after retrieving each character from the serial port hardware.  This 20 byte hook is injected over a piece of error-checking code that is not activated in the absence of errors.  The physical address and content of this error-checking code must be known to the attacker, so that the injected code can gracefully seize control and later resume normal execution.  The keylogger hook is generated from a short assembly language routine.

The main keylogger routine is 360 bytes long and sends each keystroke as a UDP packet to a hardcoded IP address.  It uses the optional netpoll API in the Linux kernel to accomplish this in such a compact payload.  This routine is generated from C code that is compiled by the attacker as though it is a part of the target kernel.  The attacker must know the addresses of the relevant netpoll routines as well as the address of a region of kernel memory that is unused, so that the keylogger can be injected into that region without interfering with the system's business functions.  We chose a region pertaining to NFS functionality.  The NFS functionality was compiled into the kernel, but is never used on this particular system.

All of the knowledge that we identified as being necessary to the attacker could reasonably be obtained if the target system is using a standard Linux distribution with a known kernel and if the attacker knows which portion of the kernel is unlikely to be used by the target system based on its purpose.  For example, other systems may use NFS, in which case it would be necessary to find a different portion of the kernel that is unused on that system in which to store the keylogger payload.

\includefig{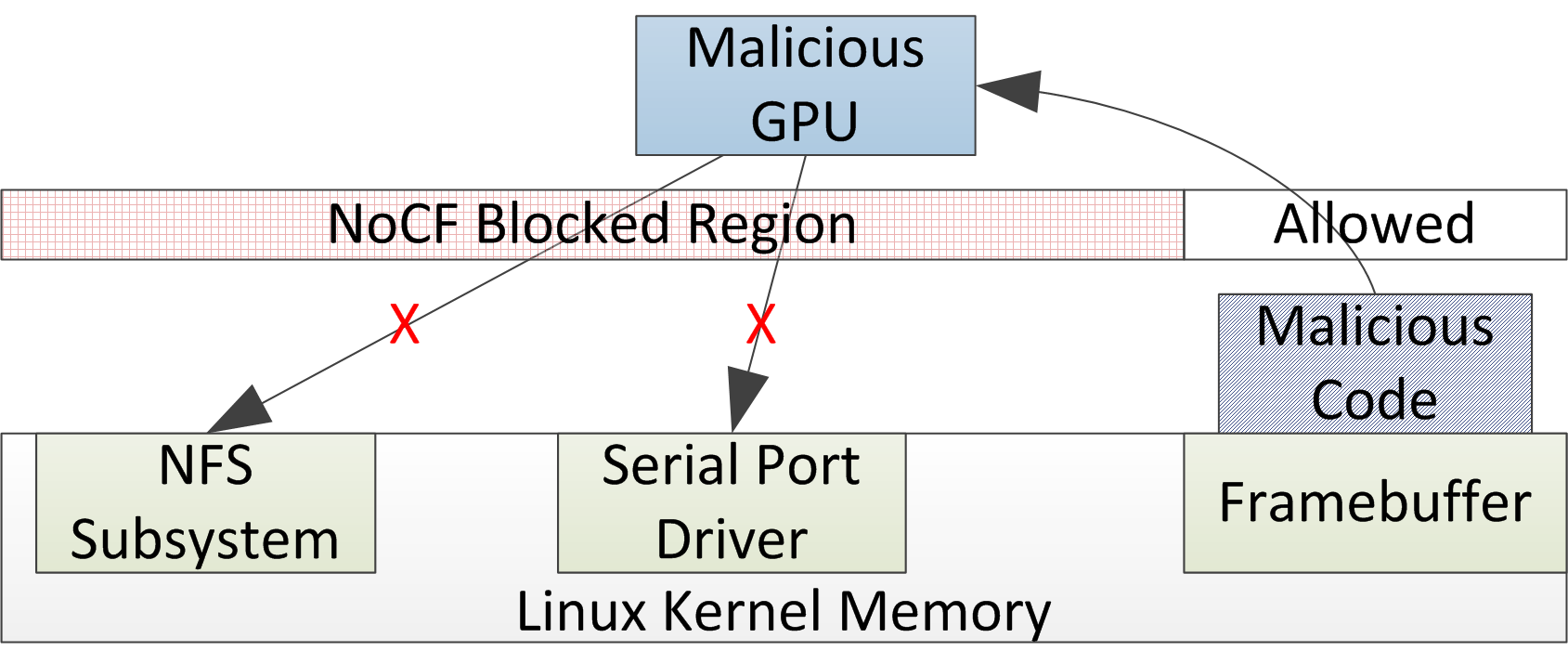}{\sys{} can be configured to block attacks that rely on writes by malicious hardware to specific memory locations that it has no legitimate need to access.}{0.7}

To constrain the GPU in such a way that this attack fails, it is simply necessary to modify the \sys{} policy to only permit accesses from the GPU to its designated framebuffer in main memory, as is depicted in Figure~\ref{fig:nocf-mal-gpu}.

This particular attack could also be blocked by a kernel integrity monitor, which ensures that only approved kernel code is permitted to execute~\cite{secvisor}.  The malware injected by the GPU would not be approved, so it would be unable to execute.  However, kernel integrity monitors fail to address attacks on userspace and can be complex, invasive, and high-overhead.

\subsection{Hardware Resource Usage}
\label{sect:eval:resources}

\sys{} requires additional hardware resources in the SoC.  The resources dedicated to the interposers should be roughly proportional to the number of NoC ports being regulated, whereas the integrity core can control many interposers, limited by the policy configuration workload imposed by those interposers.  We synthesized three distinct designs using Xilinx EDK v.13.3 to measure resource usage.  The first \textit{(Reference)} lacks all NoCF functionality, the second \textit{(Reference with Integrity Core)} adds the integrity core, and the third and fourth \textit{[\sys{} (AXI)} and \textit{\sys{} (FSL)]} add the interposers and policy configuration interconnects.  The labels in parentheses denote the type of interconnect between the interposers and integrity core.  All designs run the processor cores at 80MHz.  The FPGA hardware resource utilization is listed in Table~\ref{tbl:eval:resources}.

\begin{table*}
\centering
\begin{tabular}{p{0.36\linewidth}*{3}{p{0.1\linewidth}}p{0.16	\linewidth}} \toprule
\textbf{Design}			& \textbf{LUTs}		& \textbf{Registers}	& \textbf{Slices}	& \textbf{36Kb RAMs}
\\ \midrule
Reference			& 21,245		& 21,446		& 10,328		& 30		\\ \hline
Reference with Integrity Core	& 22,297 (5.0\%)	& 22,034 (2.7\%)	& 10,655 (3.2\%)	& 34 (13.3\%)	\\ \hline
\sys{} (AXI)			& 26,282 (23.7\%)	& 24,798 (15.6\%)	& 11,936 (15.6\%)	& 34 (13.3\%)	\\ \hline
\sys{} (FSL)			& 26,780 (26.1\%)	& 24,541 (14.4\%)	& 12,100 (17.2\%)	& 34 (13.3\%)	\\
\bottomrule
\end{tabular}
\caption{FPGA resource utilization of designs.  Percentages indicate increases over reference design.  The MicroBlaze cores in our prototype use very few hardware resources compared to cores that are popular in mobile devices, such as ARM cores, so the relative resource usage increases imposed by \sys{} in such a system will be much lower than those reported here.}
\label{tbl:eval:resources}
\end{table*}

FPGAs contain a variety of resources: \textit{1) Look-Up Tables (LUTs)}: Typically, a LUT is a small read-only memory that stores values representing the output of a logic function applied to its address input.  Some LUTs can alternately be used as small RAMs or shift registers. \textit{2) Registers}: These store values that must be saved across clock cycle transitions. \textit{3) Slices}: Each slice contains some of the LUTs and registers listed previously. \textit{4) RAMs}: These are relatively large blocks of on-chip RAM, which can be much faster than off-chip RAM such as DDR3.  An FPGA design specifies how to configure and connect all of these types of components using wires within the FPGA to implement a circuit.

\sys{} imposes negligible performance overhead in our prototype system that isolates two instances of Linux.  Due to the coarse granularity of the resource allocations to each instance, we were able to define large memory regions in the policy that resulted in a total of six delayed NoC requests during an entire Linux session involving booting the kernel, HTTP downloads, and GZip compression.  However, other workloads could result in much higher levels of overhead.  For example, a system that defines small memory regions for policy rules and then accesses many of them within a short period of time would cause \sys{} policy rules to be frequently replaced.  This effect could be reduced by increasing the number of policy rules stored in the interposers, at the cost of increased hardware resource utilization.  The cost of each interrupt that updates policy rules depends on the amount of processing performed by the interrupt handler.

\section{Formal Analysis}
\label{sect:formal}

We developed a shallow embedding of a subset of Bluespec into Maude, a native term rewriting system, and used a Maude model of \sys{} to precisely identify a subtle vulnerability in \sys{}. A shallow embedding is one where source terms are mapped to target terms whose semantics are natively provided by the target system.  We only model the portion of the system that is shown with a hatched background in Figure~\ref{fig:internals}.  This model was sufficient to detect an interesting vulnerability, although a complete model would be necessary to analyze the entire \sys{} system in the future.

We manually converted substantial portions of the Bluespec code for \sys{} to Maude using a straightforward syntactic translation method that could be automated.  We developed our Bluespec description with no special regard for its amenability to analysis, so the subset of the Bluespec syntax that we modeled has not been artificially restricted.  We modeled each variable name and each value for Bluespec structures, enumerations, and typedefs as a Maude term.  We defined separate sorts for variable names and for data values that can be placed in Bluespec registers or wires.  We defined subsorts for specific types of register data, such as the types of data that are transferred through AXI interfaces and the state values for each channel.  We defined a separate sort for policy rules.

The model was structured as an object-oriented system.  Several distinct message types can be sent between objects.  All of them specify a method to be invoked, or that was previously invoked and is now returning a value to its caller.  Anonymous and return-addressed messages are both supported.  The latter specify the originator of the message.  These are used to invoke methods that return some value.  There are staged variants of the anonymous and return-addressed message types that include a natural number indicating the stage of processing for the message.  This permits multiple rewrite rules to sequentially participate in the processing of a single logical message.  Return messages wrap some other message that was used to invoke the method that is returning.  They attach an additional piece of content, the return value, to the wrapped message.  Special read and write token messages regulate the model's execution, as will be described below.  Finally, two special types of messages are defined to model interactions over the FSL interface.  An undecided message contains an address request, modeling an interposer notifying the integrity core of a blocked request.  An enforce write message models the integrity core instructing the interposer to recheck the blocked request.  Those two message types abstract away the details of FSL communication, since those are not relevant to the Critical Security Invariant described below.

We defined equations to construct objects modeling the initial state of each part of the system.  We defined Maude object IDs as hierarchical lists of names to associate variables with the specific subsystem in which they are contained and to represent the hierarchical relationships between subsystems.  We defined five classes corresponding to the Bluespec types of variables in the model.  Registers persistently store some value.  The value that was last written into the register in some clock cycle prior to the current one is the value that can be read from the register.  A register always contains some value.  Wires can optionally store some value within a single clock cycle.  Pulse wires behave like ordinary wires, but they can only store a single unary value.  OR pulse wires behave like pulse wires, but it is possible for them to be driven multiple times within a single clock cycle.  They will only store a unary value if they are driven at least once during the clock cycle.

We modeled Bluespec methods as rewrite rules.  The required activation state for the relevant objects is written on the left hand side of the rule, and the transformed state of those objects is written on the right hand side.  Either side can contain Maude variables.  Simple Bluespec rule conditions can be represented by embedding the required variable values into the left hand side of the corresponding Maude rule.  More complex conditions can be handled by defining a conditional Maude rule that evaluates variables from the left hand side of the Maude rule.  Updates to register variables require special handling in Maude.  We define a wire to store the value to be written to the register prior to the next clock cycle, and include a Maude rewriting rule to copy that value into the register before transitioning to the next cycle.

We modeled Bluespec functions as Maude equations.  We also defined Maude functions to model complex portions of Bluespec rules, such as a conditional expression.

The main challenge that we overcame in embedding Bluespec in Maude stems from the fact that Maude by default implements something similar to pure term rewriting system semantics, in which no explicit ordering is defined over the set of rewrite rules.  To model the modified term rewriting semantics of Bluespec, we imposed an ordering on the rules in the Maude theory that correspond to Bluespec rules and restricted them to fire at most once per clock cycle.  This includes rules to model the implicit Bluespec rules that reset ephemeral state between cycles.  We used the Maude strategy framework to control rule execution~\cite{marti-oliet_rewriting_2009}.  The Bluespec compiler output a total ordering of the rules that was logically equivalent to the actual, concurrent schedule it implemented in hardware.  We applied that ordering to the corresponding Maude rules.

To model bit vectors, we relied on a theory that had already been developed as part of a project to model the semantics of Verilog in Maude~\cite{meredith_formal_2010}.

To search for vulnerabilities in \sys{}, we focused on the following Critical Security Invariant:%
\begin{invariant}
If an address request is forwarded by a \sys{} interposer, then it is permitted by the policy within that interposer.
\end{invariant}

We modeled some basic attack behaviors to search for ways in which that invariant could be violated.  In particular, we specified that during each clock cycle attackers may issue either a permissible or impermissible address request, relative to a predefined policy, or no address request.  The AMBA AXI4 specification requires master IP to wait until its requests have been acknowledged by the slave IP before modifying them in any way, but our model considers the possibility that malicious IP could violate that.

We used the Maude ``fair rewriting'' command to perform a breadth-first search of the device's possible states for violations of the Critical Security Invariant.  We extended the Maude strategy framework to trace the rule invocations so that the vulnerabilities underlying detected attacks could be independently verified and remedied.  

\includefig{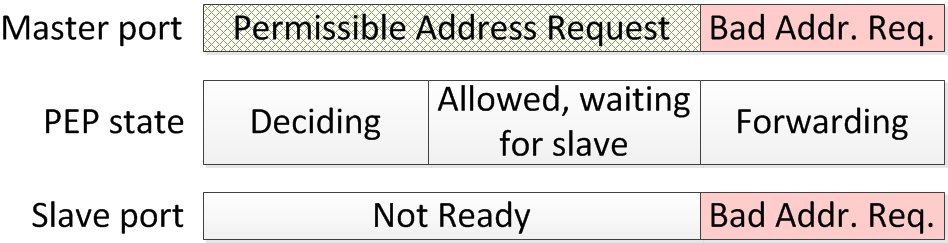}{Timing of address requests at port relative to PEP state for an attack forwarding an unchecked address request.}{0.7}

We detected a subtle possible attack applicable to a straightforward implementation of the interposer.  First, the attacker issues the permissible request, when the slave IP is not yet ready to accept a new request.  The attacker then issues the impermissible request after the PEP has approved the first request and is simply waiting for the slave IP to accept the request.  The PEP assumes the master adheres to the protocol specification and will wait for the initial request to be acknowledged, so it passes the request through from the master.  This attack is depicted in Figure~\ref{fig:wait-attack}.  This type of model is powerful, since it is a simple matter to model basic attacker behaviors which can then be automatically analyzed to detect complex attacks.

We implemented a countermeasure in the Bluespec code to block this attack.  It now buffers the request that is subjected to access control checking, and then issues that exact request to the slave if it is allowed, regardless of the current state of the request interface from the master.  We evaluated this enhanced design in \S{}\ref{sect:eval:resources}.  This countermeasure introduces additional space overhead, so it is not something that a designer would reasonably be expected to include at the outset in a straightforward implementation.

We extended the Verilog specification of the NoC ports to interface with the Verilog generated by the Bluespec compiler.  This implies that we must trust or verify the Verilog interface code.  The interface code is straightforward, consisting almost entirely of wire connections, so it should be amenable to manual or formal analysis.  Note that we must also trust the Bluespec compiler to output Verilog code that corresponds to the input Bluespec code.

Our threat model allows malicious IP to perform intra-clock cycle manipulations of the wires in the port to the interposer.  The effects of such manipulations are difficult to analyze at the level of abstraction considered in this section.  If this type of behavior is a concern, it can be easily suppressed by buffering the port using a slice of registers that effectively forces the IP to commit to a single port state for a full clock cycle from the perspective of the interposer.  The commercial NoC IP that we used in our prototype supports the creation of such register slices with a simple design parameter.  This solution would introduce a single clock cycle delay at the port.

Ultimately, \sys{} and other elements of the TCB should be formally verified to be resistant to foreseeable attack types, and the analysis described here suggests that the elegant semantics of Bluespec helps to make such an effort more tractable than it would be if we had used Verilog or VHDL.

As an intermediate goal, it will be important to model more potential attacker behaviors to identify additional vulnerabilities and formally verify the absence of vulnerabilities when possible.  The model should be expanded to model all possible sequences of values that misbehaving IP could inject into the NoC ports to which it has access.  A challenge is that each master controls a large number of input wires that feed into the NoC.  Many of these wires carry 32-bit addresses and data, so inductive proof strategies may permit that number to be substantially reduced by showing that a model using narrower address and data ports is equivalent to the full model in the context of interesting theorems.  Similarly, induction may permit long sequences of identical input values or repetitive sequences of input values to be collapsed to shorter sequences, if in fact the NoC logic does not impart significance to the patterns in question.

We considered one theorem for which we detected a counterexample, but there are many other theorems that are foundational to the system's trustworthiness and that should be used as guidance while analyzing \sys{}.  The process of identifying these theorems should be informed by past vulnerabilities, system requirements, and desirable information-theoretic properties.

Analyzing \sys{} and the rest of the TCB with respect to the theorems and the detailed model we have proposed is an important step towards providing strong assurance that the system can be trusted to process sensitive data alongside potentially misbehaving hardware and software components.  Our formal analysis of the existing \sys{} prototype demonstrates the improved analysis capabilities that are enabled by formal hardware development practices and modern formal analysis tools.  This suggests that the broader analysis effort we have proposed is feasible, given sufficient resources.

\section{Discussion}
\label{sect:discussion}

\subsection{Protecting Full-Featured Devices}
\label{sect:discussion:fine-grained}

\includefig{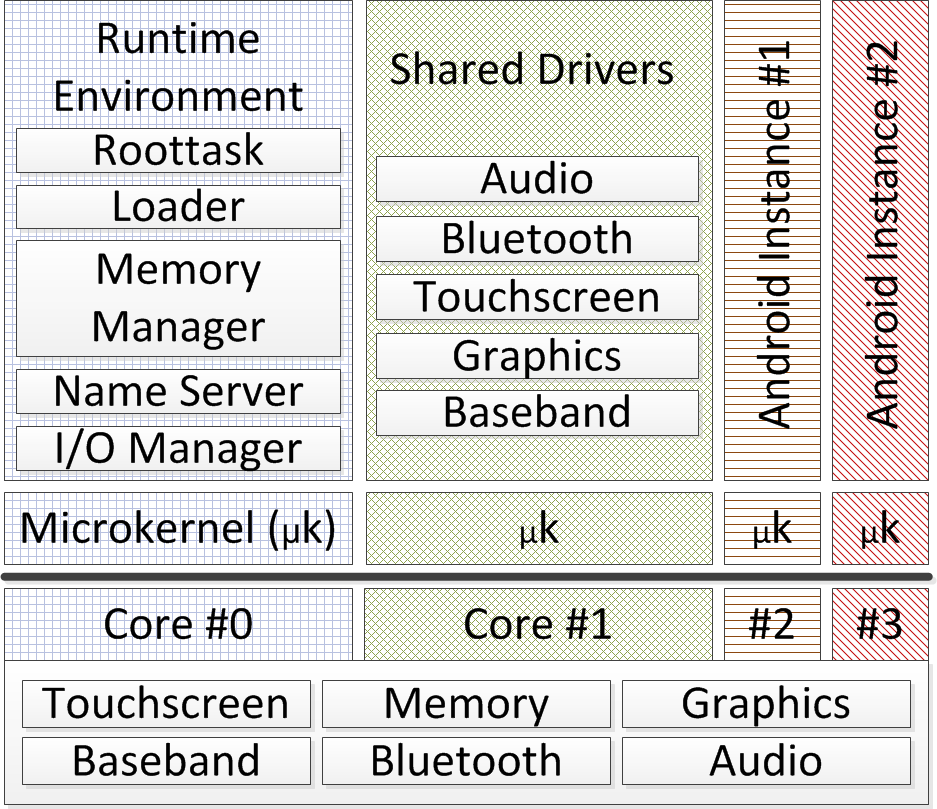}{Dual-persona smartphone architecture based on L4Android.  The thick line divides hardware below from software above.}{0.6}

In this section, we briefly explain how \sys{} can protect systems with varying degrees of sharing between protected subsystems.  We first consider a full-featured multi-persona device that runs two Android instances on a single device and, unlike our prototype, multiplexes hardware peripherals between them.  L4Android is one hypervisor that supports such a system~\cite{lange_l4android:_2011}, so we use it to provide examples of what software components are required.  We base our discussion on a quad-core processor, although additional cores could support a higher number of isolated Android instances or further subdivisions of system components.

Figure~\ref{fig:case-study-layers} depicts a possible architecture for such a system that is practical and exhibits desirable attack-resistance properties.  The grouping of software components onto particular cores is enforced by \sys{}, which allows access to the memory and peripherals associated with each group of software components only from the core beneath that group.  Some other memory regions outside of those denoted by the patterning are used to permit limited communication between software components on different cores.  Communication is implemented by a microkernel present on every core and collectively serving as the hypervisor.  Each microkernel also configures its core's MMU to isolate components sharing that core, like those shown in gray boxes.  Not shown are the integrity core and integrity kernel.  The hypervisor communicates with those components in some way to establish a policy matching the depicted boundaries and the associated shared memory regions.

The runtime environment is isolated on a core, but it provides services to instances and drivers on other cores.  Those services are fairly stable, minimal, and relatively feasible to validate through extensive testing and formal verification.

We show the shared drivers grouped on a single core, but this could permit a malicious driver for a device not needed by some instance to compromise the operation of other drivers the instance does depend upon.  This illustrates the value of dividing drivers between cores in a similar way as the instances they serve.  Historically, drivers have been unreliable for various reasons and could be particularly vulnerable to attacks.  Thus, shared drivers are a likely avenue for attacks between instances.  It is prudent to limit driver sharing as much as possible, and to focus resources on validating the remaining shared drivers.

It may be less convenient to use a mobile device with multiples instances of Android, since it is necessary to switch between them, and it is often quite useful for apps to share some data between themselves.  \sys{} can protect apps within a single-instance Android system from each other in the presence of certain CPU flaws.  For example, consider a bug in the processor's TLB that causes certain TLB flush operations to fail, such that some TLB entries incorrectly remain.  This could lead to a vulnerability if the OS swaps out memory from one process and then allocates the newly-freed physical memory to a different process.  Accesses made by the first process to the virtual memory mapped by the illegitimate TLB entries would be capable of accessing data belonging to the second process.  If we assume that those processes are on separate cores and that the OS coordinates with the integrity kernel to only allow accesses from a specific core to the memory currently allocated to code running on that core, then this vulnerability would be blocked.  We make no claims about whether this specific type of flaw is plausible, beyond noting that the TLB of a popular processor did contain a serious flaw~\cite{shimpi_amds_2008}.

\sys{} may be useful for constraining a kernel compromise within a single-instance Android system to code running on a particular core.  It would accomplish this by restricting kernel code running on a particular core to only access data that is actually relevant to its operation.

The primary challenge to implementing such a system is precisely that the kernel normally assumes it has full access to the system.  If the kernel or userspace code on a core attempts to access a resource it has not been allocated, the reliability of the system will be affected negatively.  It is necessary to fully characterize all memory accesses that may legitimately occur from every core and to concisely express those permissions as a memory access control policy.  Such a policy may need to control access to irregularly-aligned and -sized objects, which may require the use of a different policy format than the one discussed previously.  For example, Mondrian Memory Protection may be useful~\cite{witchel_mondrian_2002}.  Note that the resultant system could exhibit a combination of the best properties of microkernel and monolithic kernel systems, since parts of the kernel could be isolated from each other while all parts of the kernel would still view the entire kernel as directly-addressable space, without needing to perform explicit IPC to other portions of the kernel.  However, it is not yet clear that the accesses in a monolithic kernel can be adequately characterized to make this feasible and sufficiently restricted to substantially improve security.

Some interesting types of policies may be difficult to enforce efficiently given the current level of visibility the integrity core has into the system and the relative simplicity of the current interposer policy rule format.  A promising future direction is to mark certain memory regions as audited, so that extended information about accesses to those regions would be forwarded to the integrity core to assist it in enforcing a complex policy or monitoring those accesses.  For example, this information could include the full content about to be read from or written to the region.  As an example of how this could be useful, consider the possibility of restricting Android messages between apps.  The integrity core could monitor the application-level type information associated with the message and block specific types of messages.  This type of policy has been considered in other work, although it was implemented differently~\cite{bugiel_towards_2012}.

\subsection{Other Applications}
\label{sect:discussion:other-apps}

Desktop and server systems may have similar problems with untrusted IP as mobile devices, but their system topologies are substantially different and would require different technical solutions to those problems.  For example, a desktop CPU may communicate over wires exposed on a motherboard to a chipset that provides access to peripherals, and over a different set of wires to memory chips.  Interposers would be required at each location where a bus master connected to an interconnect, which would necessitate a different \sys{} topology and protocols.  For example, error detection support would be required in the protocol to handle errors introduced by transmitting data over wires between chips.  Thus, while this paper may be instructive for solutions on other types of platforms, we focus on the unique characteristics of mobile devices.

Much of \sys{} is specific to AMBA AXI4, but it could be ported to other types of interconnects.  To maintain maximum performance and to minimize the trusted computing base, \sys{} should be thoroughly re-engineered to match the unique characteristics of each type of interconnect.

We have already discussed the importance of protecting medical data on mobile devices, but another important application for SoCs in this domain is in the construction of specialized medical devices themselves~\cite{amit_nanda_adopting_2009}.

A possible vulnerability in such devices could originate with the wireless interface logic, since wireless is being used to communicate with an increasing number and variety of devices.  For example, the wireless driver in a blood glucose monitor may contain a buffer overflow that could be exploited by some malicious display device to install malware.  That malware could cause the device to report incorrect readings to other display devices that the individual uses as a source of data to be input into an insulin delivery device.  \sys{} could be configured to isolate the wireless driver on a dedicated core and prevent the malware from infecting the targeted functions of the device outside of the wireless subsystem.

The position of the interposers would make them well-suited for enforcing policies that regulate other aspects of accesses besides their locations.  For example, DoS attacks could be a concern within a system.  The interposers could monitor the bandwidth and other characteristics of communication flows that it forwards and block those that are excessive.

\subsection{Attestation}

The stakeholders interested in each isolated software component may require assurances that their software is properly protected.  Such assurances can be provided in the form of an attestation from \sys{} that it is running a trusted integrity kernel that enforces an acceptable policy.  There are a variety of remote attestation systems that could be used to accomplish that task~\cite{coker_principles_2011}.

An alternative approach is to enforce a whitelist, so that only integrity kernel code on the whitelist is permitted to run at all.  Seshadri \etal{} demonstrated such an approach in software for kernel code~\cite{secvisor}.  LeMay and Gunter demonstrated how to implement code whitelisting in an efficient manner using judicious hardware extensions~\cite{lemay_enforcing_2011}.

\section{Related Work}
\label{sect:related}

We consider two classes of related work: \textit{1)} tools and techniques that enable formal reasoning about hardware and \textit{2)} efforts to enhance security protections for SoC/NoC using hardware design.  The primary novelty of \sys{} is that it is a NoC access control mechanism designed to be amenable to formal analysis.

\subsection{High-Assurance Hardware Development}
\label{sect:related:ha-hw}

Advances have been made in languages for formally specifying information-flow properties in hardware like Caisson~\cite{li_caisson:_2011}.  Tiwari \etal{} developed and verified an information-flow secure processor and microkernel, but that was not in the context of a mobile-phone SoC and involved radical modifications to the processor compared to those required by NoC-based security mechanisms~\cite{tiwari_crafting_2011}.  Volpano proposed dividing memory accesses in time to limit covert channels~\cite{volpano_towards_2008}.  Information-flow techniques could be generally applicable to help verify the security of the trusted components identified in \S{}\ref{sect:design:threat}.

Other techniques are complementary to these lines of advancement in that they offer approaches for satisfying the assumptions of our threat model.  ``Moats and Drawbridges'' is the name of a technique for physically isolating components of an FPGA and connecting them through constrained interfaces so that they can be analyzed independently~\cite{HuffmireBWSKLNI08,huffmire_handbook_2010}.

SurfNoC schedules multiple protection domains onto NoC resources in such a way that non-interference between the domains can be verified at the gate level~\cite{wassel_surfnoc:_2013}.  This could complement \sys{} by preventing unauthorized communications channels between domains from being constructed in the NoC fabric.

Richards and Lester defined a shallow, monadic embedding of a subset of Bluespec into PVS and performed demonstrative proofs using the PVS theorem prover on a 50-line Bluespec design~\cite{richards_monadic_2011}.  Their techniques may be complementary to our model checking approach for proving properties that are amenable to theorem proving.  Katelman defined a deep embedding of BTRS into Maude~\cite{michael_kahn_katelman_meta-language_2011}.  BTRS is an intermediate language used by the Bluespec compiler.  Our shallow embedding has the potential for higher performance, since we translate Bluespec rules into native Maude rules.  Bluespec compilers could potentially output multiple BTRS representations for a single design, complicating verification.  Finally, our embedding corresponds more closely to Bluespec code, which could make it easier to understand and respond to output from the verification tools.

\subsection{SoC/NoC Protection}
\label{sect:related:soc-prot}

The line of work about NoC access control started with the Security Enhanced Communication Architecture (SECA)~\cite{coburn_seca:_2005}, a mechanism for monitoring and controlling bus data transfers using a variety of address- and value-based stateful and stateless policies.  Those policies provide a high level of expressiveness at the cost of increased complexity compared to \sys{}, which would complicate formal analysis.  SECA is validated using a multi-core mobile phone SoC containing a cryptoprocessor that shares a RAM with the other cores and requires protection for its portion of the RAM.

Cotret \etal{} extended distributed address-based controls with support for encrypting and monitoring the integrity of data sent to memory outside the SoC~\cite{cotret_distributed_2011,cotret_bus-based_2012}.  These mechanisms can address more powerful threats, but introduce additional area overhead and latency.

Some NoC-based approaches use the main NoC to carry policy configuration traffic, which is efficient and flexible~\cite{fiorin_secure_2008}.  One design uses dedicated virtual channels on a shared NoC~\cite{diguet_noc-centric_2007}.  \sys{} dedicates physically-separate interconnects between the integrity core and interposers, making it simpler to determine that only an authorized entity, the integrity core, can specify the policy.

NoC-MPU involves an MPU controlling each master's access to the NoC~\cite{porquet_noc-mpu:_2011}.  Each MPU uses policies stored in NoC-accessible memory as page tables and cached in permission lookaside buffers.  The policies are parameterized on memory addresses and compartment identifiers.  Compartment identifiers can distinguish software on a single physical master device.  The authors envision a ``global trusted agent'' running on a dedicated processor to configure page tables, which is analogous to the \sys{} integrity kernel.  Supporting in-memory page tables increases the complexity of each MPU compared to \sys{} interposers.

Kim and Villasenor focused on the threat of Trojan IP and implemented measures to prevent illicit transfers that rely on the broadcast nature of a particular bus type~\cite{kim_system--chip_2011}.  A contribution of their paper that is complementary to \sys{} is their approach for dealing with availability attacks launched by malicious IP.  They also briefly mention the implications of protocol violations by Trojan IP, but do not analyze those in depth.

Other works have proposed encryption and integrity monitoring for sensitive code and data in main memory~\cite{mccune_cpu_2011,champagne_scalable_2010}.  These confidentiality and integrity enforcement techniques do not address access control for peripherals and any unencrypted data, so it is still important to implement such access control in a verifiable manner.

Some NoC protection mechanisms have been developed commercially, but the details of their designs and the analysis processes that have been applied to them are not publicly available.  Texas Instruments has applied for a patent on hardware technology that includes the ability to restrict bus accesses and execute operating systems in isolation on virtual cores~\cite{conti_virtual}.

ARM TrustZone defines secure and non-secure modes for virtual processor cores and other bus masters~\cite{trustzone}.  Accesses from a master in non-secure mode may be restricted by an interconnect or slave devices.  Some aspects of divisions between virtual cores are implemented internally by the physical core hosting them, improving resource sharing but also increasing complexity compared to the approach in \sys{} of concentrating the PEP at the interconnect level to divide physical cores.  \sys{} also permits all bus masters to be regulated independently, whereas TrustZone permits a secure virtual core to access all system resources.  However, TrustZone does allow certain other bus masters to be restricted based on their identity.  TrustZone also defines other types of SoC security features.

Huffmire \etal{} proposed a mechanism to convert memory access policies into Verilog that could be synthesized into a reference monitor~\cite{huffmire_policy-driven_2006}.  They targeted reconfigurable systems in which the reference monitor could be replaced at runtime by reconfiguring a portion of the FPGA.  Such systems are not yet commonplace.

Individual cores or groups of cores on a Tilera Tile processor are capable of running independent instances of Linux, and cores can be partitioned using hardware access control mechanisms that block individual links between cores~\cite{tilera}.  However, those mechanisms are not applied to memory controller and cache links, which are regulated using TLBs.  \sys{} could potentially be used to regulate such links.

\section{Conclusion}
\label{sect:conclusion}

Mobile devices that became popular for personal use are increasingly being relied upon to process sensitive data, but they are not sufficiently trustworthy to make such reliance prudent.  Various software-based techniques are being developed to process data with different levels of sensitivity in a trustworthy manner, but they assume that the underlying hardware memory access control mechanisms are trustworthy.  We discuss how to validate this assumption by introducing a NoC Firewall that is amenable to formal analysis.  We present a prototype \sys{} that is implemented using a hardware description language with elegant semantics.  We demonstrate its utility by using it to completely isolate two Linux instances without running any hypervisor code on the cores hosting the instances, and to block attacks from a malicious GPU.

\section*{Acknowledgments}

This work was supported by HHS 90TR0003-01 (SHARPS) and NSF 13-30491 (ThaW). The views expressed are those of the authors only.  We measured lines of code using David A. Wheeler's 'SLOCCount'.

A revised version of this paper was published in \underline{Logic, Rewriting, and Concurrency}. The final publication is available at \url{http://link.springer.com/chapter/10.1007/978-3-319-23165-5_19}.

\bibliographystyle{abbrv}
\bibliography{arxiv-1}

\appendix

\section{Integrity Kernel}
\label{app:integrity-kernel}

This appendix provides pseudocode listings for the integrity kernel.  It focuses on the version of the prototype that uses a dedicated FSL link for each interposer, but there are only minor differences between that version and the version based on a shared AXI interconnect.

A key difference between this integrity kernel and one based on MMU or IO-MMU protections is that it does not need to configure memory protections for itself.  The integrity kernel is installed in a dedicated region of memory that is inaccessible from the other cores in the system.

The integrity kernel only implements functionality to configure memory protections, resulting in a very small TCB.  Microkernels are alternative pieces of software that can implement memory protections, but they sometimes implement many other functions that result in a larger TCB, such as inter-process communication and scheduling.  Monolithic kernels typically have even larger TCBs.

\mathsymb{Intrs}
\mathsymb{HandleIntr}

\widepseudocode{int-kern-main}{Main entrypoint.}{
\Procedure{Main}{}
  \State $\Call{IntrControllerInit}{}$ \Comment{Library call to initialize interrupt controller.}
  \ForAll{$i \in \msIntrs$} \Comment{$\msIntrs$ is a global variable with information about each interrupt line.  Each interposer has a dedicated interrupt line.}
    \State $\Call{IntrControllerSetup}{i, \msHandleIntr}$ \Comment{Library calls to configure interrupt controller to cause a specific interrupt signal to invoke a shared interrupt handler routine.}
  \EndFor
  \State $\Call{IntrControllerStart}{}$ \Comment{Start handling interrupts.}
  \While{$\top$} \Comment{Wait forever, handling interrupts.  Kernel never exits, except when the system is powered down.}
  \EndWhile
\EndProcedure
}

\mathsymb{intrId}
\mathsymb{intrData}
\mathsymb{regionSize}
\mathsymb{baseAddress}
\mathsymb{isReadAccess}

\widepseudocode{int-kern-hnd-intr}{Interrupt handler.}{
\Procedure{HandleIntr}{intrId}
  \State $\msintrData \gets \Call{ReceiveIntr}{\msintrId}$ \Comment{Retrieve information about the blocked access that caused the interrupt.}
  \State $\msregionSize \gets \Call{CalculateRegionSize}{\msintrData}$ \Comment{Calculate size of region of memory that should be granted to accessing master IP.  This routine is implemented as a collection of case statements that take as parameters the identity of the master IP and the memory region being accessed.}
  \If{$\msregionSize \ne 0$} \Comment{A region size of zero signifies a denied access.}
    \State $\Call{Grant}{\msintrId, \msintrData.\msbaseAddress, \msregionSize}$ \Comment{$\msintrData.\msbaseAddress$ specifies the base address of the region of memory being accessed.}
  \EndIf
  \State $\Call{Enforce}{\msintrId, \msintrData.\msisReadAccess}$
  \State $\Call{IntrControllerAckIntr}{\msintrId}$ \Comment{Library call to acknowledge handling the interrupt.}
\EndProcedure
}

\mathsymb{command}
\mathsymb{NewRule}

\widepseudocode{int-kern-grant}{Insert new policy rule to grant access to a memory region.}{
\Procedure{Grant}{intrId, baseAddress, regionSize}
  \State $\mscommand \gets \Call{BuildCommand}{}$
  \State $\,\,\,(\msNewRule, baseAddress, regionSize)$ \Comment{$\Call{BuildCommand}{}$ is actually implemented as a simple series of bitwise operations to construct a 32-bit command to be sent to the interposer.}
  \State $\Call{SendCommand}{\msintrId, \mscommand}$
\EndProcedure
}

\widepseudocode{int-kern-send-cmd}{Send command to interposer.}{
\Procedure{SendCommand}{intrId, command}
  \State $\Call{FSLSendCommand}{\mscommand, \msintrId}$ \Comment{Library call to send command data over FSL interconnect.}
\EndProcedure
}

\mathsymb{raw}

\widepseudocode{int-kern-recv-intr}{Retrieve data about interrupt from interposer.}{
\Procedure{ReceiveIntr}{intrId}
  \State $\msraw \gets \Call{FSLReceiveCommand}{\msintrId}$ \Comment{Library call to read command data over FSL interconnect.}
  \State \Return{$\Call{DecodeIntrData}{\msraw}$} \Comment{Decode the 32-bit representation of the interrupt cause into a structure that can be easily processed.}
\EndProcedure
}

\mathsymb{isRead}
\mathsymb{EnforceRead}
\mathsymb{EnforceWrite}

\widepseudocode{int-kern-enforce}{Command the interposer to re-evaluate the blocked access relative to the updated access control rules.}{
\Procedure{Enforce}{intrId, isRead} \Comment{$\msisRead$ should be set to true if the blocked access is a read, and false if it is a write.}
  \If{$\msisRead$}
    \State $\mscommand \gets \Call{BuildCommand}{\msEnforceRead}$
  \Else
    \State $\mscommand \gets \Call{BuildCommand}{\msEnforceWrite}$
  \EndIf
  \State $\Call{SendCommand}{\msintrId, \mscommand}$
\EndProcedure
}

\section{\sys{} Hardware}
\label{app:nocf-hardware}

This appendix presents details of the Bluespec hardware design for the components described in this paper.

We defined two types of parameterized basic interfaces, one each for unbuffered inputs and outputs.  They are parameterized on the type of data that is passed through the interface. The interface that represents inputs into the interposer declares two methods.  One allows the hardware communicating with the interposer to transfer a unit of data.  The other indicates whether the interposer is ready to receive input on the interface.  The methods for the output interface are similar.  One allows the interposer to transfer a unit of data to hardware outside the interposer, and the other allows that hardware to indicate to the interposer when data can be sent over the interface.

We aggregate one each of unbuffered input and output interfaces that transfer AXI address requests to form an address filter interface.  The address filter interface also defines two additional methods that are used internally within the interposer.  The first returns the current address request being processed, and the second is used to inform the address filter of an access control decision for the current request.

The address filter is always in one of three states: idle, committed, or waiting.  It is initially in the idle state.  Its state also includes information on the current address request being issued by the regulated AXI master, the address request that was most recently committed (see section~\ref{sect:formal} for a discussion of why it is necessary to commit each address request), and three pulse wires.  A Bluespec pulse wire is used for signaling between methods within a single clock cycle.  The pulse wires indicate whether the committed address request is valid, whether the address request was forwarded in the current clock cycle, and whether the current address request is compliant with policy, respectively.

A filter must be in the idle state to accept a new address request.  It will remain in the idle state if the NoC fabric is ready to receive the request and if any of the policy rules stored in the interposer allow the request, since the request can be passed through the interposer during that same cycle in such a case.  If any of the policy rules stored in the interposer allow the request, but the NoC fabric is not yet ready, then the filter transitions to the waiting state.  Otherwise, it enters the committed state.  The request is buffered during both the waiting and committed states.

A filter in the committed state waits for an access control decision from the integrity core.  A negative decision will cause the filter to return to the idle state without transferring the buffered request to the NoC fabric.  A positive decision will cause a transition to the idle state if the NoC fabric is ready to receive the request, and to the waiting state otherwise.

A filter in the waiting state simply waits for the NoC fabric to become ready to accept the buffered request.  It then transitions back to the idle state.

A collection of unbuffered interfaces are aggregated to form a single master or slave AXI interface.  All of the unbuffered interfaces in the master interface are directly connected to the corresponding interfaces in the slave interface, with the following exceptions.  First, the read and write response channels are temporarily taken over by the interposer when the interposer is sending a response due to a blocked request.  The interposer also notifies the connected slave device that it is not ready to receive response data during such an event.  Finally, each of the address request channels is routed through an address filter. 

The interposer sends each faulting address to the integrity core along with a bit indicating whether a read or write was attempted to or from that address.  The integrity core can respond with a specification for a new rule that the interposer should install.  It comprises a partial address, two bits indicating whether reads and/or writes are permitted, and four bits defining the size of the covered address range.

The FSL master interface enqueues an outgoing transfer to a FIFO between the interposer and the integrity core when the FIFO has available storage space, blocking when the FIFO is full.  The FSL slave interface accepts incoming transfers from the FIFO.  

The top-level interposer interface aggregates one each of an AXI master and slave interface, and one each of an FSL master and slave interface.

The main \sys{} module can hold a configurable number of rules at any point in time.  The number is set at the time the hardware is defined.  By default, none of the policy rules allow anything.

Both of the read and write channels can each be in one of the following states: permit, enforce, request, wait, check, respond, or resume.  The permit state is not actually not ordinarily used, but we added dormant support for it so that it can be used for debugging.  Each channel has an address filter interface.

The relative age of each policy rule is tracked, so that the oldest one can be replaced when necessary.

Both directions of the FSL interface are buffered using FIFOs.

We defined a function to check each policy rule against each address request on either the read or write channel, with an argument to indicate whether a read or write request is being evaluated.  The function compares the relevant portion of the addresses in the request and the policy rule (depending on the region size defined in the policy rule) and returns whether the rule allows the request if it matches.  We defined a second function to invoke the former function on all policy rules in parallel and return true if any rule allows the request.

For each channel, the current request (if any) is checked against the policy and the decision is made available to other parts of the interposer that need it.  There are two possible exceptions to this.  First, no decisions are issued when the policy is currently being updated.  Second, an allow decision is issued whenever the policy is not being updated and the channel is currently in the permit state.

When a channel is in the permit or enforce states and an allow decision was issued in the current cycle as described above, then that decision is sent to the address filter for that channel.  The address filter's response to that decision is described above.

If a channel is in the enforce state and a deny decision was issued in the current cycle, then that channel transitions to the request state.

During the single cycle in which a channel is the request state, it enqueues an FSL transfer with information about the AXI request that is pending, including whether it is a read or write request and the address.  The channel then transitions to the wait state.  The channel remains in the wait state until the integrity core sends a packet specifically indicating that the channel should transition to the check state.

A channel in the check state forwards the decision derived from the updated policy to the address filter interface and then transitions to the resume state for an allow decision or the respond state for a deny decision.  It also stores the ID for the current AXI request.  In addition to that, the read channel also stores the length of the read request.

A channel in the resume state transitions to the enforce state in the next cycle.  The resume state could be eliminated from the design to slightly reduce latency without any degradation in functionality.

A channel in the respond state sends a decode error response back to the master device, which is the same type of error that would be sent if a missing device had been addressed.  The write channel then transitions to the enforce state.  The read channel maintains the decode error response for the number of cycles required by the AXI protocol and then transitions to the resume state.

\end{document}